\documentclass[numberedappendix,12pt,preprint]{emulateapj}

\usepackage[]{amsmath}

\shorttitle{EoR \lya\ from UV bright galaxies}
\shortauthors{Mason et al. (2018)}

\usepackage{color}			   
\definecolor{midgray}{gray}{0.4}		
\definecolor{orange}{rgb}{1,0.5,0}  
\usepackage[shortlabels]{enumitem}

\usepackage[colorlinks=true,citecolor=blue,linkcolor=blue]{hyperref}    

\usepackage{etoolbox}
\makeatletter
\patchcmd{\NAT@citex}
  {\@citea\NAT@hyper@{\NAT@nmfmt{\NAT@nm}\NAT@date}}
  {\@citea\NAT@nmfmt{\NAT@nm}\NAT@hyper@{\NAT@date}}
  {}
  {}
\patchcmd{\NAT@citex}
  {\@citea\NAT@hyper@{%
     \NAT@nmfmt{\NAT@nm}%
     \hyper@natlinkbreak{\NAT@aysep\NAT@spacechar}{\@citeb\@extra@b@citeb}%
     \NAT@date}}
  {\@citea\NAT@nmfmt{\NAT@nm}%
   \NAT@aysep\NAT@spacechar%
   \NAT@hyper@{\NAT@date}}
  {}
  {}
\patchcmd{\NAT@citex}
  {\@citea\NAT@hyper@{%
     \NAT@nmfmt{\NAT@nm}%
     \hyper@natlinkbreak{\NAT@spacechar\NAT@@open\if*#1*\else#1\NAT@spacechar\fi}%
       {\@citeb\@extra@b@citeb}%
     \NAT@date}}
  {\@citea\NAT@nmfmt{\NAT@nm}%
   \NAT@spacechar\NAT@@open\if*#1*\else#1\NAT@spacechar\fi%
   \NAT@hyper@{\NAT@date}}
  {}
  {}
\makeatother

\newcommand{\simgt}{\,\rlap{\lower 3.5 pt \hbox{$\mathchar \sim$}} \raise
1pt \hbox {$>$}\,}
\newcommand{\simlt}{\,\rlap{\lower 3.5 pt \hbox{$\mathchar \sim$}} \raise
1pt \hbox {$<$}\,}

\newcommand{\msun}{M_{\odot}}

\newcommand{\lya}{Ly$\alpha$}
\newcommand{\pEW}{$p(EW_{\mathrm{Ly}\alpha})$}
\newcommand{\pEWem}{$p(EW_{\mathrm{Ly}\alpha}^\mathrm{emit})$}
\newcommand{\pEWobs}{$p(EW_{\mathrm{Ly}\alpha}^\mathrm{obs})$}

\newcommand{\xHI}{{\overline{x}_\textsc{hi}}}
\newcommand{\DV}{{\Delta v}}
\newcommand{\Tigm}{{\mathcal{T}_\textsc{igm}}}
\newcommand{\MUV}{M_\textsc{uv}}

\newcommand{\HST}{\textit{HST}}
\newcommand{\JWST}{\textit{JWST}}
\newcommand{\WFIRST}{\textit{WFIRST}}

\newcommand{\BE}{\begin{equation}}
\newcommand{\EE}{\end{equation}}
\newcommand{\BEA}{\begin{eqnarray}}
\newcommand{\EEA}{\end{eqnarray}}

\defcitealias{Mason2018}{M18}	

\begin{document}

\title{Beacons into the Cosmic Dark Ages:\\ Boosted transmission of \lya\ from UV bright galaxies at $z \simgt 7$}

\author{
Charlotte A. Mason$^{1}$,
Tommaso Treu$^{1}$,
Stephane de Barros$^{2,3}$,
Mark Dijkstra$^{4}$,
Adriano Fontana$^{5}$,\\
Andrei Mesinger$^{6}$,
Laura Pentericci$^{5}$,
Michele Trenti$^{7,8}$, and
Eros Vanzella$^{3}$,
}
\affil{$^{1}$ Department of Physics and Astronomy, UCLA, Los Angeles, CA, 90095-1547, USA}
\affil{$^{2}$ Observatoire de Gen\`{e}ve, Universit\`{e} de Gen\`{e}ve, 51 Ch. des Maillettes, 1290 Versoix, Switzerland}
\affil{$^{3}$ INAF Osservatorio Astronomico di Bologna, via Gobetti 93/3, 40129 Bologna, Italy}
\affil{$^{4}$ Institute of Theoretical Astrophysics, University of Oslo, P.O. Box 1029, N-0315 Oslo, Norway}
\affil{$^{5}$ INAF Osservatorio Astronomico di Roma, Via Frascati 33, I-00040 Monteporzio (RM), Italy}
\affil{$^{6}$ Scuola Normale Superiore, Piazza dei Cavalieri 7, I-56126 Pisa, Italy}
\affil{$^{7}$ School of Physics, University of Melbourne, Parkville, Victoria, Australia}
\affil{$^{8}$ ARC Centre of Excellence for All Sky Astrophysics in 3 Dimensions (ASTRO 3D)}
\email{cmason@astro.ucla.edu}

\begin{abstract}
Recent detections of Lyman alpha (\lya) emission from $z>7.5$ galaxies were somewhat unexpected given a dearth of previous non-detections in this era when the intergalactic medium (IGM) is still highly neutral. But these detections were from UV bright galaxies, which preferentially live in overdensities which reionize early, and have significantly Doppler-shifted \lya\ line profiles emerging from their interstellar media (ISM), making them less affected by the global IGM state. Using a combination of reionization simulations and empirical ISM models we show, as a result of these two effects, UV bright galaxies in overdensities have $>2\times$ higher transmission through the $z\sim7$ IGM than typical field galaxies, and this boosted transmission is enhanced as the neutral fraction increases. The boosted transmission is not sufficient to explain the observed high \lya\ fraction of $\MUV \simlt -22$ galaxies \citep{Stark2017}, suggesting \lya\ emitted by these galaxies must be stronger than expected due to enhanced production and/or selection effects. Despite the bias of UV bright galaxies to reside in overdensities we show \lya\ observations of such galaxies can accurately measure the global neutral hydrogen fraction, particularly when \lya\ from UV faint galaxies is extinguished, making them ideal candidates for spectroscopic follow-up into the cosmic Dark Ages.
\end{abstract}

 
\section{Introduction}
\label{sec:intro}

Reionization of hydrogen in the universe's first billion years was driven by the first sources of light. Accurately measuring the timeline of reionization, i.e. average neutral hydrogen fraction ($\xHI$) as a function of redshift, enables us to infer properties of these sources. \lya\ emission from galaxies has long been touted as a tracer of $\xHI$ during reionization: \lya\ photons are absorbed by neutral hydrogen \citep[e.g.,][]{Dijkstra2014a}. 

The rapidly declining fraction of Lyman-break galaxies (LBGs) emitting \lya\ at $z>6$ \citep[e.g.,][]{Fontana2010a,Stark2010,Treu2013,Schenker2014,Pentericci2014,Mason2018} and strong damping wing absorption of $z\sim7$ quasar spectra  \citep{Greig2016b,Banados2017} suggest the universe is significantly neutral at $z\simgt7$. Recent detections of \lya\ from galaxies at $z>7.5$ \citep{Roberts-Borsani2015,Zitrin2015a,Oesch2015,Stark2017} are therefore surprising. Furthermore, these detections come from $\MUV \simlt -22$ galaxies ($\simgt 2.5 L^*$). At lower redshifts UV bright galaxies are least likely to have strong \lya\ \citep[e.g.][]{Stark2010}. Why can we see \lya\ from these galaxies?

Reionization is likely highly inhomogeneous -- overdense regions reionize more rapidly as they are filled with many ionizing sources \citep[e.g.,][]{McQuinn2007a}. The brightest galaxies likely reside in overdensities \citep[e.g.,][]{Trenti2012,Barone-Nugent2014,Castellano2016a}. How easily are \lya\ photons from such galaxies in overdensities transmitted through the IGM, compared to field galaxies? How does the ISM radiative transfer of \lya\ affect its IGM transmission? (Figure~\ref{fig:cartoon}) Can these biased galaxies still measure $\xHI$?

Here we combine cosmological reionization simulations with empirical models of galaxy properties to understand the transmission of \lya\ from UV bright galaxies. We describe our combination of simulations and empirical models in Section~\ref{sec:method}. In Section~\ref{sec:res} we present our results on the evolving transmission of \lya\ emission from galaxies in massive halos, the interpretation of the observed `\lya\ fraction', and the efficacy of UV bright galaxies as probes of $\xHI$. We discuss our results in Section~\ref{sec:dis} and summarize in Section~\ref{sec:conc}.

We use the \citet{PlanckCollaboration2015} cosmology. All magnitudes are in the AB system.

\section{Method}
\label{sec:method}

\begin{figure*}[t!] 
\centering
\includegraphics[width=0.9\textwidth]{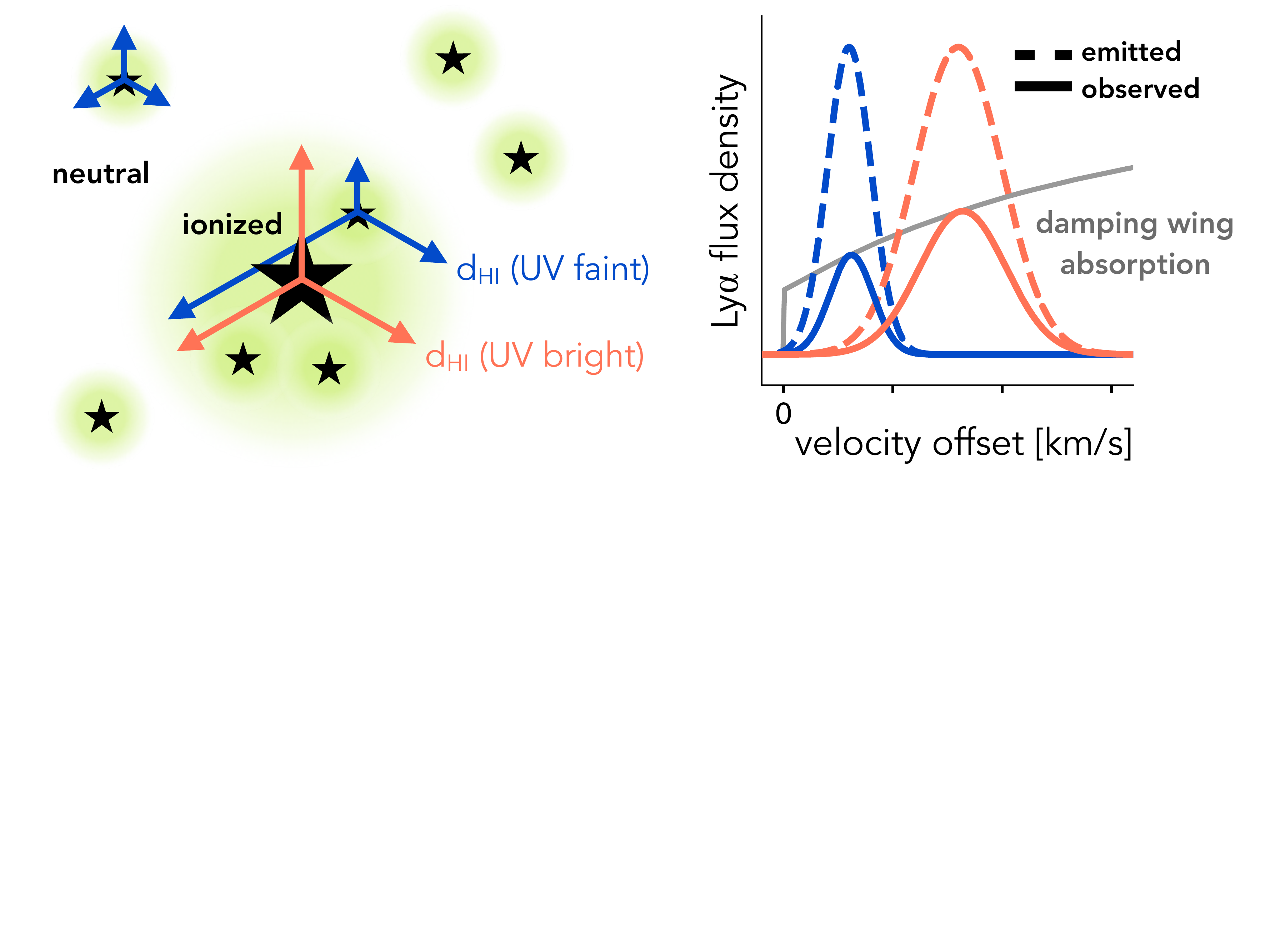}
\caption{\textbf{Left:} UV bright galaxies (large stars) preferentially live in overdensities, which reionize early (green regions). \lya\ damping wing optical depths are dominated by the distance to the first neutral patch (white regions) photons encounter, $d_\textsc{hi}$, thus UV bright galaxies have higher average \lya\ transmission than UV faint galaxies (small stars) as they live further from neutral patches. The sightline-to-sightline scatter of \lya\ transmissions for UV bright galaxies is lower due to lower scatter in $d_\textsc{hi}$ (orange arrows for UV bright galaxies, blue arrows for UV faint galaxies). \textbf{Right:} Gas and dust resonantly scatter and absorb \lya\ photons inside galaxies. \lya\ emitted (dashed lines) by UV bright galaxies (orange) is usually more Doppler-shifted than \lya\ from UV faint galaxies (blue), as they contain more gas and dust. Damping wing absorption during reionization attenuates \lya\ smoothly with wavelength/velocity offset (gray line -- example shown is $\xHI=0.66$), so \lya\ transmitted (solid lines) through the IGM depends on galaxy properties.}
\label{fig:cartoon}
\end{figure*}

To model the transmission of \lya\ photons from galaxies through the reionizing IGM we combine the public Evolution of Structure simulations \citep[EoS\footnote{\url{http://homepage.sns.it/mesinger/EOS.html}},][]{Mesinger2016} with empirical models of galaxy properties. We follow the method of \citet{Mason2018} \citepalias[hereafter][]{Mason2018} and refer the reader there for more details. We describe our methods briefly below.

The EoS simulations treat inhomogeneous recombinations and ionizations at a sub-grid level on a density field in a $1024^3$ box with sides 1.6 Gpc. The simulations have two runs to bracket the potential reionization parameter space: \textsc{Faint Galaxies}, where ionization sources are primarily low mass galaxies, producing reionization morphologies characterized by small HII patches; and \textsc{Bright Galaxies}, where reionization is dominated by more massive galaxies, producing larger HII patches. We use the fiducial \textsc{Faint Galaxies}, but show in Section~\ref{sec:res_transmission} our results do not significantly depend on the choice of simulation run. 

We populate dark matter halos in the simulations with physically motivated galaxy properties: UV luminosities from the \citet{Mason2015a} model; and emitted \lya\ rest-frame equivalent widths (EW) and \lya\ line velocity offsets from source galaxies' systemic redshifts, $\DV$. We use the empirical model presented by \citetalias{Mason2018} where $\DV$ is correlated with halo mass to encompass the complex ISM radiative transfer. Massive halos have higher $\DV$, likely due to increased scattering in their denser ISM (Figure~\ref{fig:cartoon}, right panel). We model lines as Gaussians, centered at $\DV$, with $FWHM = \DV$(for $\DV < 20$ km s$^{-1}$ we set $FWHM = 20$ km s$^{-1}$, comparable to the maximum expected thermal broadening). We assume the observed $z\sim6$ \lya\ EW distribution, \pEW, is equivalent to the emitted $z\sim7$ \pEWem\ (i.e. the change observed between these redshifts is due to reionization only) and use a $p(EW_{\mathrm{Ly}\alpha}|\MUV)$ fit to the $z\sim6$ sample presented by \citet{DeBarros2017} \citepalias[the fit is described by][and accounts for \lya\ non-detections]{Mason2018}.

A key quantity we compute is the differential \textit{transmission fraction} of \lya\ photons through the IGM: $\Tigm(\xHI) = EW(\xHI)/EW(\xHI = 0)$. We calculate $\Tigm$ by modeling the emitted \lya\ lineshape and attenuating it with the damping wing absorption optical depth ($\tau_\textsc{igm}$) from cosmic neutral hydrogen patches along the line of sight in the simulations:
\BE \label{eqn:Tfromtau}
\Tigm(\xHI, M_h, \DV) = \int_0^\infty \mathrm{d}v \; J_\alpha(\DV, M_h, v) e^{-\tau_\textsc{igm}(\xHI, M_h, v)}
\EE
where $J_\alpha(\DV, M_h, v)$ is the normalized \lya\ lineshape emitted from galaxies. We model circumgalactic medium (CGM) absorption by truncating the lineprofiles at the halo circular velocity. As we are only interested in the differential evolution of EW this is valid assuming the only significant change in the optical depth to \lya\ between $z\sim6$ and $z\sim7$ is due to reionization. We discuss the impact of an evolving CGM in Section~\ref{sec:dis}.

We calculate $\Tigm$ for millions of realizations of model galaxies along thousands of sightlines in 40 $z=7$ IGM simulation cubes with average neutral fractions $0 \leq \xHI \leq 0.95$ ($\Delta \xHI \sim 0.02$) to generate $p(\Tigm | \xHI, M_h)$ and forward-model the observed \pEW.

\section{Results}
\label{sec:res}

Here we describe the key results of our study: \lya\ from UV bright galaxies in massive halos can have high transmission through the IGM, even in a highly neutral universe (Section~\ref{sec:res_transmission}); our model is consistent with the observed evolution of the \lya\ fraction, except for extremely bright galaxies ($\MUV \simlt -22$) which must have higher than expected emitted \lya\ EWs (Section~\ref{sec:res_forecast}); and UV bright galaxies can measure $\xHI$ if their emitted \lya\ EW distribution is known (Section~\ref{sec:res_reionization}).

\subsection{Boosted transmission of \lya\ from massive halos}
\label{sec:res_transmission}

\begin{figure}[!t] 
\centering
\includegraphics[width=0.49\textwidth]{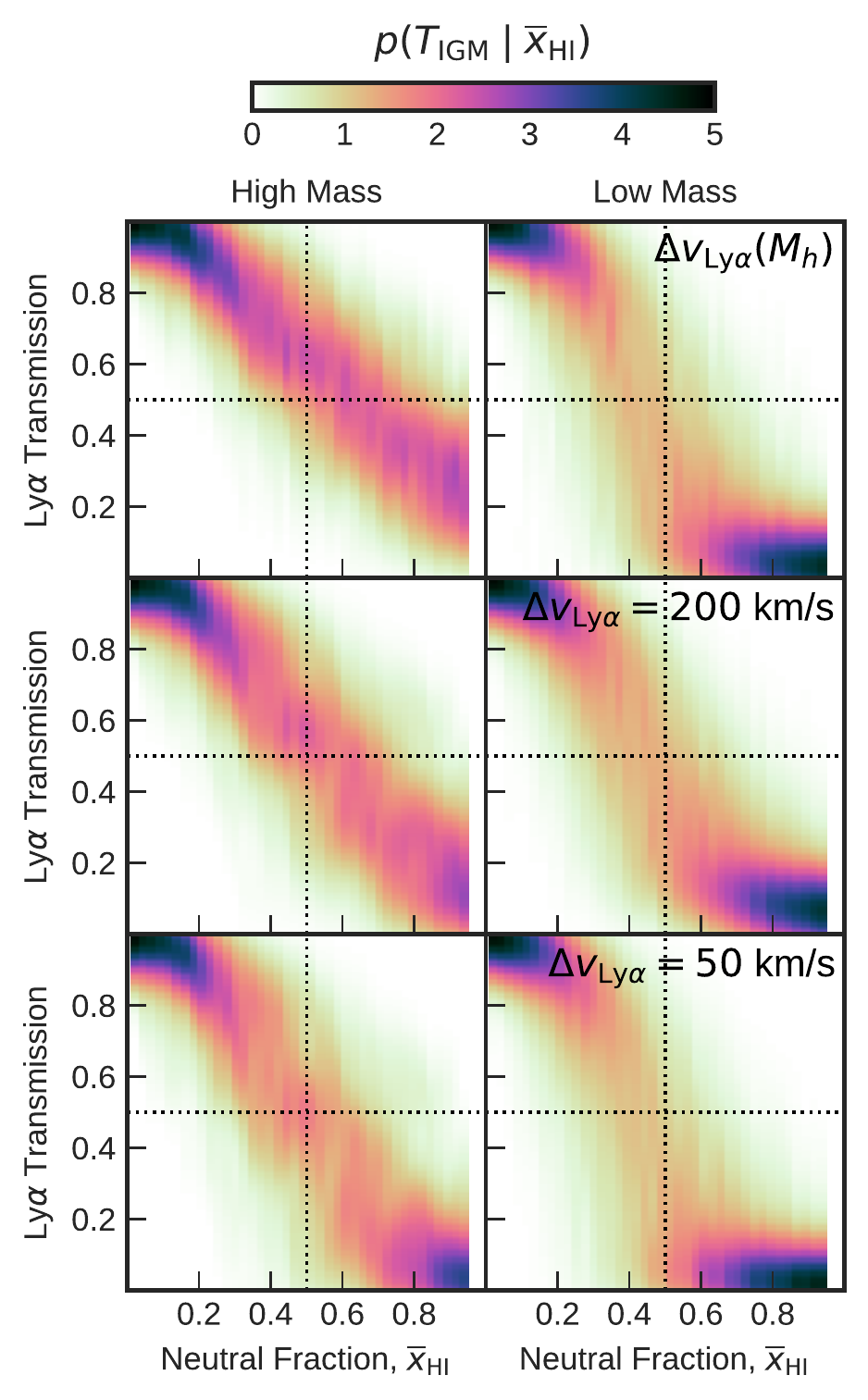}
\caption{\lya\ transmission fraction distributions, $p(\Tigm)$ at a given $\xHI$. Highest density/darkest regions correspond to most likely values of $\Tigm$ at each $\xHI$. We show $p(\Tigm)$ in two mass bins ($10^{11.5} \leq M_h \leq 10^{12}\msun$, left; $10^{10} \leq M_h \leq 10^{11}\msun$, right). We use three models for emitted \lya\ lines: (upper panels) the mass-dependent model presented by \citetalias{Mason2018}, with high mass halos having higher $\DV$; (middle) $\DV=200$ km s$^{-1}$; (lower) $\DV=50$ km s$^{-1}$. With mass-dependent velocity offsets $\Tigm$ is boosted for high mass halos.}
\label{fig:T}
\end{figure}

\begin{figure}[!t] 
\centering
\includegraphics[width=0.46\textwidth]{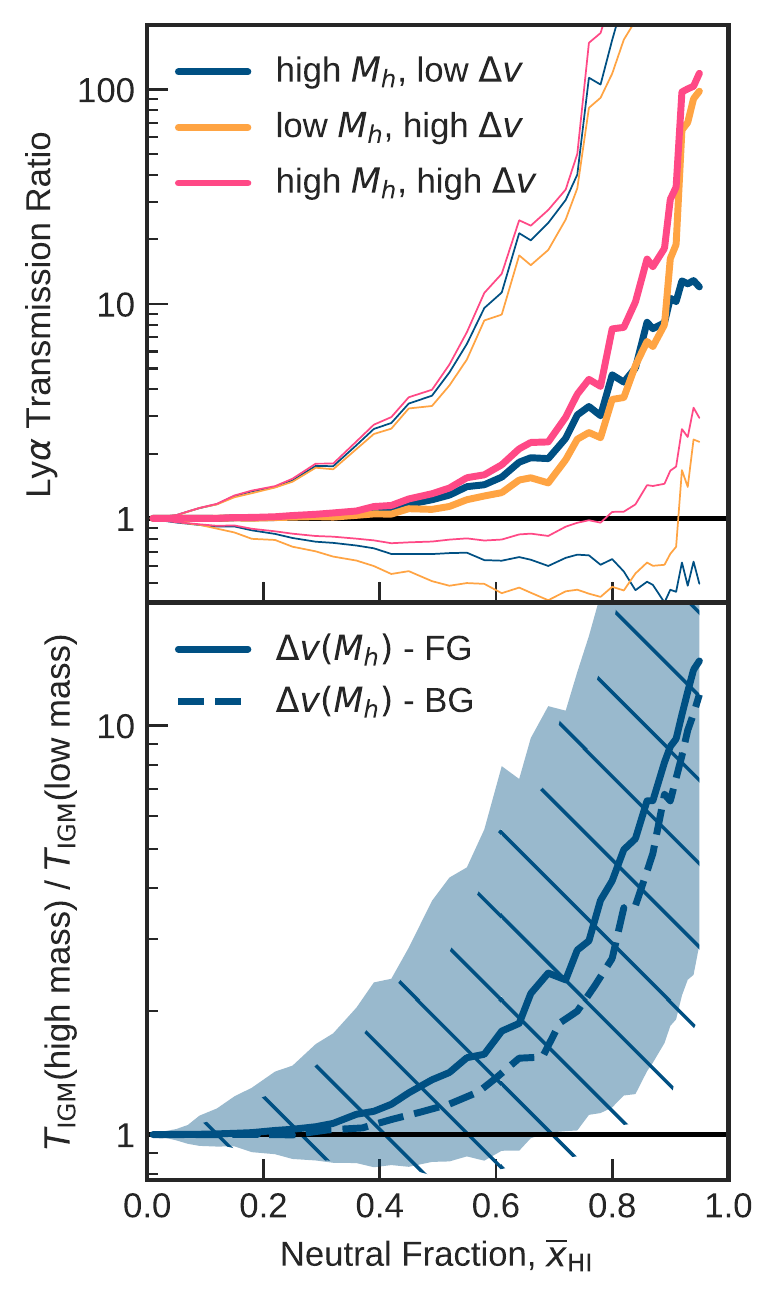}
\caption{\textbf{Upper:} Ratio of $\Tigm$ comparing galaxies in low mass halos with $\DV = 50$ km s$^{-1}$ with galaxies in: high mass halos, $\DV = 50$ km s$^{-1}$ (blue); low mass halos, $\DV = 200$ km s$^{-1}$ (orange); high mass halos, $\DV = 200$ km s$^{-1}$ (pink). We use $p(\Tigm)$ and mass bins from Figure~\ref{fig:T}. We plot median ratios as solid thick lines and $16-84\%$ range (due to sightline-to-sightline variations) as thin solid lines. The biggest boost is for galaxies in massive halos with high $\DV$. \textbf{Lower:} Ratio of $\Tigm$ for galaxies in low and high mass bins, assuming mass-dependent $\DV$. We show the ratio derived using the two EoS simulation runs: the fiducial \textsc{Faint Galaxies} (median - solid line, $16-84\%$ - shaded region); \textsc{Bright Galaxies} (median - dashed line, $16-84\%$ - hashed region). $\Tigm$ boosting in massive halos is relatively insensitive to simulation choice.}
\label{fig:T_med}
\end{figure}

\begin{figure*}[t] 
\centering
\includegraphics[width=0.99\textwidth]{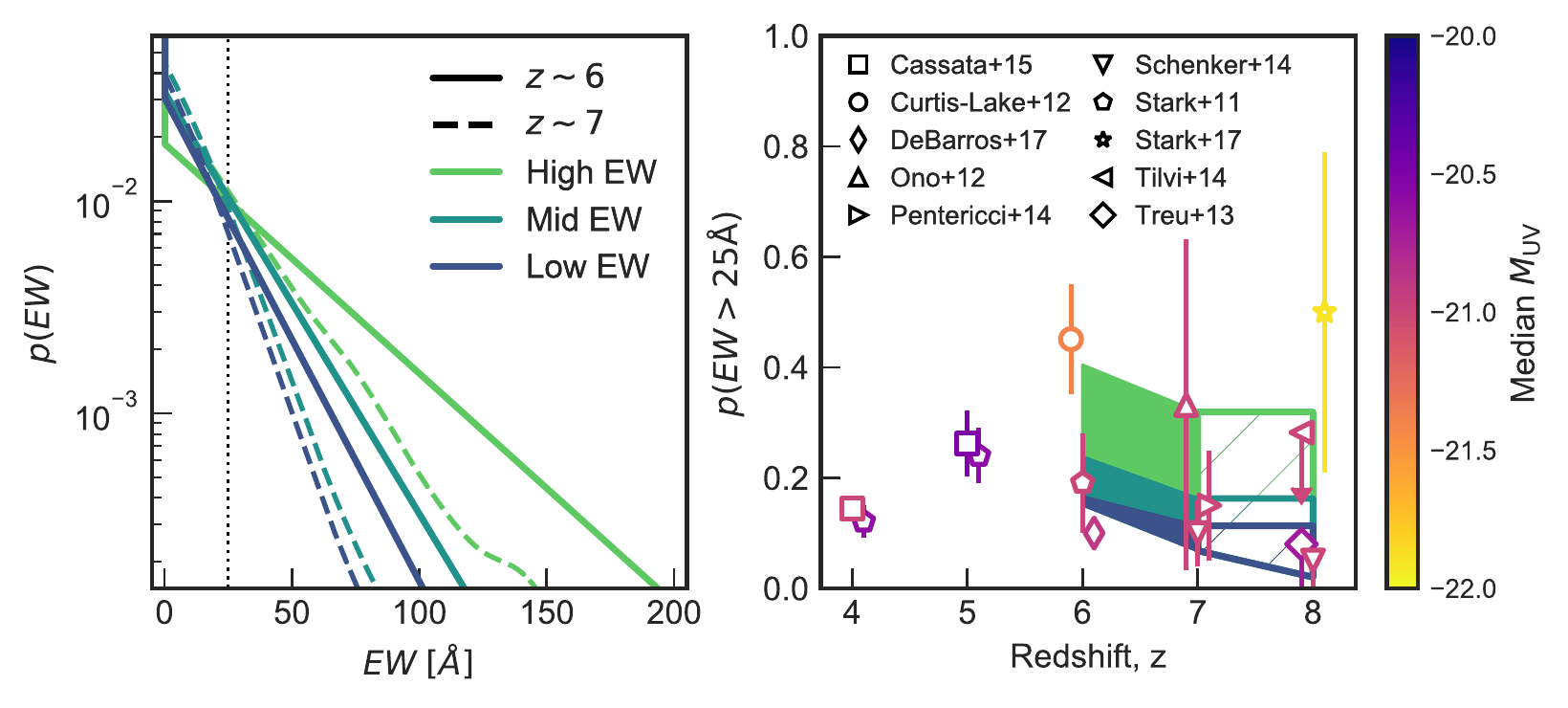}
\caption{Redshift evolution of \pEW\ for galaxies in massive halos, for three model \pEW\ (Low, Mid and High). \textbf{Left:} Each \pEW\ model at $z\sim6$ \citep[solid, fit from][]{DeBarros2017} and $z\sim7$ \citepalias[dashed, assuming the inferred median $\xHI=0.59$ by][]{Mason2018}. \textbf{Right:} The fraction of LBGs showing \lya\ $EW > 25$\AA\ (dotted black line in left panel). We plot observations of UV bright samples \citep[indicated by point shape,][]{Stark2011,Curtis-Lake2012,Ono2012,Treu2013,Tilvi2014,Pentericci2014,Schenker2014,Cassata2015,Stark2017,DeBarros2017}, with the samples' median $\MUV$ indicated by color. Shaded regions show the range of evolution allowed by the \citetalias{Mason2018} $\xHI$ constraints ($16-84\%$ range), for a given model \pEW. Hashed regions indicate the allowed evolution to $z=8$ assuming $\xHI$ does not increase.}
\label{fig:forecast}
\end{figure*}

To explore the differences between the most biased galaxies and the bulk of the high redshift galaxy population we examine $p(\Tigm)$ in two halo mass bins: $10^{11.5} \leq M_h \leq 10^{12} \msun$ (57 sightlines in the EoS simulations, hosting $\MUV \simlt -21$ galaxies) and $10^{10} \leq M_h \leq 10^{11}$ \citep[$\sim10^4$ sightlines, $\MUV \simgt -19.5$ galaxies,comparable to the faintest $z>6$ LBGs with detected \lya, e.g.][]{Huang2016,Hoag2017}.

Figure~\ref{fig:T} shows $p(\Tigm | \xHI, M_h)$, using three models for \lya\ velocity offsets: (1) drawn from the \citetalias{Mason2018} $p(\DV | M_h)$ model: low mass halos have median $\DV \sim90$ km~s$^{-1}$, high mass halos have median $\DV\sim220$ km~s$^{-1}$. (2) $\DV=200$ km~s$^{-1}$, often the fiducial value used in reionization \lya\ modeling \citep[e.g.,][]{Dijkstra2011,Mesinger2014}. (3) $\DV=50$ km~s$^{-1}$.

Irrespective of emitted line properties, galaxies in massive halos have higher $\Tigm$, as they preferentially live in overdensities which reionize early \citep[e.g.,][]{McQuinn2007a}, so their \lya\ photons are redshifted into the flattest part of the damping wing (Figure~\ref{fig:cartoon}) by the time they reach cosmic neutral patches. Mass-dependent velocity offsets enhance this effect: \lya\ from low mass halos is more easily attenuated as they have low $\DV$, whereas $\Tigm$ from massive halos is boosted.

The scatter in $\Tigm$ is lower for massive halos: the smaller scatter in distance from source galaxies to the first neutral patch (Figure~\ref{fig:cartoon}) reduces the sightline-to-sightline variation in optical depths. This makes galaxies in massive halos accurate probes of $\xHI$. The effect is most pronounced for $0.3 \simlt \xHI \simlt 0.6$, when neutral patches are narrower and more widely separated \citep{Mesinger2008}. As noted by \citet{Mesinger2008} (though in the context of quasars), if halo masses can be estimated for galaxies the accuracy in $\xHI$ increases.

The top panel of Figure~\ref{fig:T_med} investigates contributions to $\Tigm$. We compare $\Tigm$ from galaxies in low and high mass halos, with fixed low or high $\DV$, to galaxies in low mass halos with low $\DV$. Massive halos always have high $\Tigm$, as they reside in larger ionized bubbles, indicating halo mass is the dominant cause of high transmission. When \lya\ is emitted at high $\DV$ $\Tigm$ is significantly boosted for massive halos. In the very early stages of reionization $\Tigm$ is boosted for low mass halos with high $\DV$ compared to low $\DV$, massive halos, likely because ionized bubbles around massive halos are still small.

The lower panel of Figure~\ref{fig:T_med} shows a realistic estimate of the boosting, using mass-dependent $\DV$ (comparing the top panels of Figure~\ref{fig:T}). For $\xHI > 0.6$ $\Tigm$ for massive halos are $>2\times$ higher than for low mass halos, rising to a factor $\sim10$ for $\xHI > 0.9$. We compare the transmission ratio for the two EoS simulations: \textsc{Faint Galaxies} and \textsc{Bright Galaxies} (Section~\ref{sec:method}). The transmission boost is comparable; this effect is relatively independent of the timeline and morphology of reionization. We use these realistic $\Tigm$ for UV bright galaxies in the next sections.


\subsection{Evolving \lya\ fraction for UV bright galaxies}
\label{sec:res_forecast}

An increasing fraction of \lya\ emitters (EW $> 25$\AA) is observed in the LBG population over $2 \simlt z \simlt 6$ \citep[e.g.,][]{Stark2010,Cassata2015}, likely due to decreasing dust in galaxies \citep{Hayes2011}. A drop in the \lya\ fraction at $z>6$ is usually attributed to absorption by an increasingly neutral IGM during reionization \citep[see][for a recent review]{Dijkstra2014a}.

Figure~\ref{fig:forecast} (right panel) shows the $4 \leq z \leq 8$ \lya\ fraction for UV bright galaxies. At $z<6$ the observations are consistent, but at $z \geq 6$ the \lya\ fraction measured for samples with $\MUV \simlt -21.5$ \citep{Curtis-Lake2012,Stark2017} is significantly higher than for those at lower luminosities. Much of this discrepancy may be due to selection effects: using only the $z_{850}$-band for LBG selection the \citet{Curtis-Lake2012} sample could be biased towards strong \lya\ emission \citep{DeBarros2017}, and the \citet{Stark2017} sample was selected via red Spitzer/IRAC [3.6]-[4.5] colors \citep{Roberts-Borsani2015} making them likely strong [O\textsc{III}]+H$\beta$ emitters, requiring hard radiation fields and young stellar populations, which increase \lya\ production and escape \citep{Finkelstein2013,Zitrin2015a}. Using our model we test how the boosted $\Tigm$ for galaxies in massive halos (Section~\ref{sec:res_transmission}) contributes to their \lya\ emitter fraction.

We plot the evolution allowed by the \citetalias{Mason2018} $z\sim7$ neutral fraction estimate ($\xHI = 0.59_{-0.15}^{+0.11}$) for galaxies in massive halos, using the maximum transmission demonstrated in Figure~\ref{fig:T} (top left panel). We forward-model \pEWobs\ by convolving $p(\Tigm)$ with the UV magnitude-dependent \pEWem\ described in Section~\ref{sec:method}. \pEWem\ is a major uncertainty so we use a range of distributions: \textsc{Low-EW}, \textsc{Mid-EW} and \textsc{High-EW}, corresponding to the measured $z\sim6$ distributions for LBGs with $\MUV \sim \{-21, -20.5, -20\}$, which bracket the EW variation in the \citet{DeBarros2017} sample (Figure~\ref{fig:forecast}, left panel). Based on $z\leq6$ observations we expect UV bright galaxies to have \textsc{Low-EW} or \textsc{Mid-EW} distributions.

The observed evolution of the \lya\ fraction for $\MUV > -21.5$ samples is consistent with negligible evolution in \pEWem. The \textsc{High-EW} distribution is required to be consistent with the \citet{Stark2017} \lya\ fraction error region, which is unexpected given UV bright galaxies at lower redshifts tend to have low \lya\ EWs \citep[e.g.,][]{Stark2010}.


\subsection{UV bright galaxies as probes of reionization}
\label{sec:res_reionization}

To test the efficacy of UV bright galaxies as probes of reionization we perform a Bayesian inference to obtain the posterior distribution of the neutral fraction given simulated observations of galaxies with \lya\ EW and $\MUV$ measurements: $p(\xHI \; | \; \{EW_{\mathrm{Ly}\alpha}, \MUV \})$. Using Bayes' theorem this posterior is proportional to $\prod_i p(EW_{\mathrm{Ly}\alpha,i} \; | \; \xHI, M_{\textsc{uv}A,i}) \times p(\xHI)$, assuming the observations are independent. 

We follow the method described by \citetalias{Mason2018} and generate the likelihood of observing a given EW: $p(EW_{\mathrm{Ly}\alpha} \; | \; \xHI, \MUV)$, by convolving the high halo mass $p(\Tigm)$, described in Section~\ref{sec:res_transmission}, with a distribution of emitted EWs. For this work we consider $\MUV = -22$ galaxies. By assigning galaxies to a range of halo masses we include some scatter in $M_h - \MUV$ \citep[e.g.,][]{Finkelstein2015b}. To generate mock observations we draw EW values from a likelihood for a given $\xHI$, convolve with a 5\AA\ uncertainty and treat galaxies with EW $<15$\AA\ as non-detections (which are robustly accounted for in the inference). We perform the inference using these mock observations. 

\begin{figure}[t] 
\centering
\includegraphics[width=0.49\textwidth]{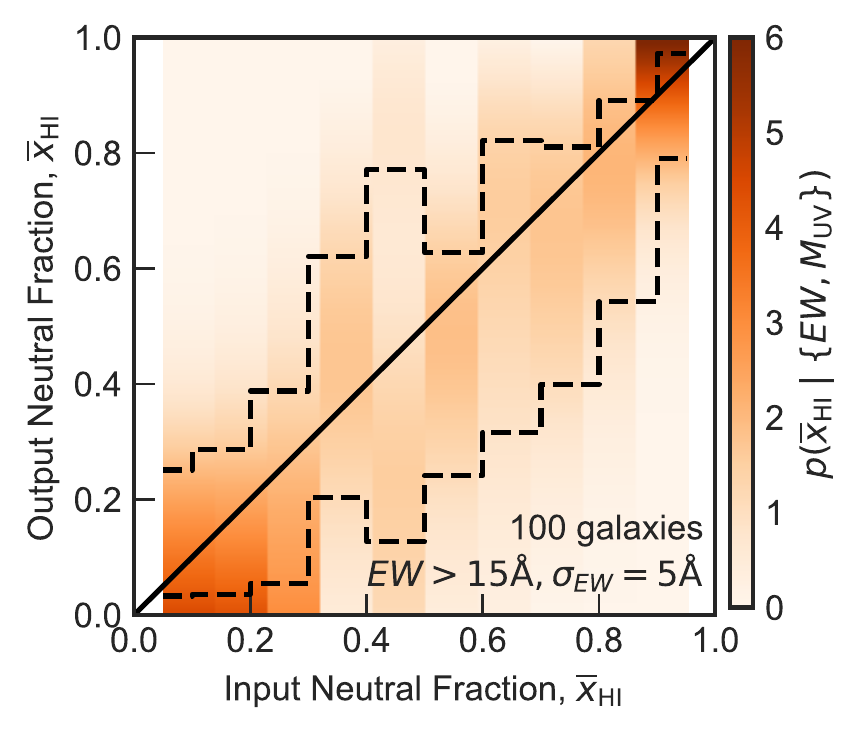}
\caption{Posterior distributions for $\xHI$ for a grid of input $\xHI$ using simulated samples of 100 $\MUV = -22$ galaxies. Each vertical strip is a posterior for input $\xHI$, darkest colors show the highest probabilities. The one-to-one relation between input and output $\xHI$ is shown as a solid black line. We plot the $16-84\%$ range for each posterior (within dashed lines). The posteriors are consistent with the input $\xHI$ within this range.}
\label{fig:posterior_test}
\end{figure}

Figure~\ref{fig:posterior_test} shows the posteriors obtained using a simulated sample of 100 UV bright galaxies for a grid of input $\xHI$ values. The inferred posteriors are consistent with the input values over the entire range within the $16-84\%$ region, showing UV bright galaxies can be accurate tracers of the average IGM state. We note our posteriors are broad ($\Delta \xHI \sim 0.4$) compared to those obtained using fainter galaxies \citepalias[c.f. $\Delta \xHI \sim 0.25$ in][]{Mason2018}. This uncertainty is driven by the shape ƒof \pEW, which declines with increasing EW. Reionization further kills the high EW tail of the distribution, making high EW objects rare.\\
\newline

\section{Discussion}
\label{sec:dis}

We have shown UV bright galaxies in high mass halos can be precise probes of reionization and are increasingly valuable in reionization's early stages when \lya\ in UV faint galaxies, emitted close to systemic velocity, is overwhelmingly absorbed in the IGM. However, there are two limitations to using such galaxies to probe reionization: (1) they are rare; (2) they emit less \lya\ due to absorption in their dense ISM. Below we discuss prospects for overcoming these limitations.

Wide-area photometric surveys such as the Brightest of Reionizing Galaxies survey \citep[BoRG,][]{Trenti2011}, UltraVISTA and UDS \citep[e.g.,][]{Bowler2015} and GOLDRUSH \citep{Ono2017} have discovered $\sim100$ $z\simgt7$ $\MUV \simlt -21$ LBGs in $\sim100$ deg$^2$. Future wide-area surveys with e.g., \WFIRST\ \citep{Spergel2015} and \textit{Euclid} \citep{Laureijs2011} will likely increase this by a factor $\simgt100$ in $>15,000$ deg$^2$. These sources will be ideal candidates for spectroscopic follow-up to measure the \lya\ EW distribution needed to infer the neutral fraction.

Do UV bright galaxies emit less \lya? Whilst most $z\simlt6$ observations indicate they do \citep[e.g.,][]{Verhamme2008,Stark2010}, our results suggest the $z\sim8$ galaxies presented by \citet{Stark2017} must have high intrinsic \lya\ EW. Recent observations of a $\MUV \sim -22$ galaxy after reionization at $z\sim4$ detected \lya\ emission with low $\DV$ and Lyman continuum radiation \citep{Vanzella2017b}, suggesting significantly ionized pathways through the ISM and/or CGM from such galaxies. If these galaxies are efficient producers of ionizing radiation, they may also increase their local ionization field to boost \lya\ transmission through the CGM/IGM.

A holistic understanding of \lya\ emission as a function of redshift and galaxy properties is therefore crucial to improve the use of \lya\ as a cosmological tool. These measurements are becoming increasingly feasible with multi-wavelength observations of LBGs, and time should be invested in establishing \lya\ emission properties over a wide redshift and galaxy mass/UV magnitude range, both in wide areas, and in deep lensed fields with \HST\ \citep[e.g.,][]{Treu2015,Schmidt2016} and in the near future with \JWST\ \citep{2017jwst.prop.1324T}. Better measurements of these properties will enable us to disentangle IGM, CGM and ISM effects. 

\section{Summary and Conclusions}
\label{sec:conc}

We have investigated the IGM transmission of \lya\ from UV bright galaxies during the Epoch of Reionization by combining reionization simulations and empirical relations for galaxy and \lya\ properties. Our main conclusions are:

\begin{enumerate}[(i)]
\item \lya\ emitted by UV bright galaxies in massive halos has a higher mean and lower dispersion in IGM transmission than \lya\ from typical field galaxies in low mass halos. This is primarily due to massive halos predominantly residing in overdensities which reionize early, and boosted by their higher \lya\ velocity offsets, reducing damping wing absorption by cosmic neutral hydrogen.
\item This boosted transmission is not sufficient to explain the observed evolution of the $6 \simlt z \simlt 8$ \lya\ fraction for extremely UV bright galaxies \citep{Stark2017}, suggesting these objects have higher emitted \lya\ EW than expected.
\item With sufficient numbers, the observed \lya\ EW distribution of UV bright galaxies can place tight constraints on the IGM neutral fraction during reionization, and may be the only way to probe the IGM at $z>7$ when quasars are exceedingly rare and \lya\ from most UV faint galaxies is extinguished.
\end{enumerate}

More comprehensive measurements of the \lya\ EW distribution as a function of redshift and galaxy properties are necessary to understand the evolving visibility of \lya\ emission and to disentangle the effects of the ISM and IGM during reionization. Current and upcoming spectroscopic observations have the ability to this and increase the efficacy of \lya\ as a cosmological tool.

\acknowledgments
We thank Dan Stark and Crystal Martin for useful discussions. CM acknowledges support through the NASA Earth and Space Science Fellowship Program Grant NNX16AO85H. AM acknowledges European Research Council support under the European Union's Horizon 2020 research and innovation program (grant No 638809 - AIDA). MT acknowledges support by the Australian Research Council (awards FT130101593 and CE170100013). This work was supported by \HST\ BoRG grants GO-12572, 12905, 13767 and 15212, and \HST\ GLASS grant GO-13459. 

\textit{Software:} IPython \citep{Perez2007}, matplotlib \citep{Hunter2007}, NumPy \citep{VanderWalt2011}, and EMCEE \citep{Foreman-Mackey2013}.


\end{document}